\documentstyle[floats,aps]{revtex}
\parskip=\medskipamount
  \font\elevenmib=cmmib10 scaled 1095
  \font\tenmib=cmmib10
  \font\eightmib=cmmib10 scaled 800
  \font\sixmib=cmmib10 scaled 667
  \skewchar\elevenmib='177
  \newfam\mibfam
  
  \textfont\mibfam=\tenmib
  \scriptfont\mibfam=\eightmib
  \scriptscriptfont\mibfam=\sixmib
  \mathchardef\alpha="710B
  \mathchardef\beta="710C
  \mathchardef\gamma="710D
  \mathchardef\delta="710E
  \mathchardef\epsilon="710F
  \mathchardef\zeta="7110
  \mathchardef\eta="7111
  \mathchardef\theta="7112
  \mathchardef\kappa="7114
  \mathchardef\lambda="7115
  \mathchardef\mu="7116
  \mathchardef\nu="7117
  \mathchardef\xi="7118
  \mathchardef\pi="7119
  \mathchardef\rho="711A
  \mathchardef\sigma="711B
  \mathchardef\tau="711C
  \mathchardef\phi="711E
  \mathchardef\chi="711F
  \mathchardef\psi="7120
  \mathchardef\omega="7121
  \mathchardef\varepsilon="7122
  \mathchardef\vartheta="7123
  \mathchardef\varrho="7125
  \mathchardef\varphi="7127

\def\etal{{\it et al.\/}}

\def\ie{{\it i.e.\/}}

\def\eg{{\it e.g.\/}}

\def\sss#1{{\scriptscriptstyle #1}}

\def\ssr#1{{\sss{\rm #1}}}

\def\kT{k_{\sss\rmB}T}

\def\dsl{\raise.15ex\hbox{$/$}\kern-.57em\hbox{$\partial$}}
\def\nsl{\raise.15ex\hbox{$/$}\kern-.57em\hbox{$\nabla$}}
\def\id{\raise.72ex\hbox{$-$}\kern-.85em\hbox{$d$}\,}
\def\gtwid{\,{\raise.3ex\hbox{$>$\kern-.75em
\lower1ex\hbox{$\sim$}}}\,}
\def\ltwid{\,{\raise.3ex\hbox{$<$\kern-.75em
\lower1ex\hbox{$\sim$}}}\,}
\def\undr{\raise.3ex\hbox{$\sim$\kern-.75em\lower1ex\hbox{$|\vec
x|\to\infty$}}}

\def\frac#1#2{{\textstyle{#1 \over #2}}}
\def\half{\frac{1}{2}}
\def\third{\frac{1}{3}}

\def\({\left (}
\def\){\right )}
\def\cbl{\left\{}
\def\cbr{\right\}}

\def\cC{{\cal C}}

\def\cH{{\cal H}}

\def\cO{{\cal O}}

\def\xhi{{\raise.35ex\hbox{$\chi$}}}

\def\rmA{{\rm A}}
\def\rmB{{\rm B}}

\def\rmD{{\rm D}}

\def\rmG{{\rm G}}

\def\rmT{{\rm T}}

\def\rmV{{\rm V}}

\def\rmc{{\rm c}}

\def\rmm{{\rm m}}
\def\rmn{{\rm n}}

\def\rms{{\rm s}}
\def\rmt{{\rm t}}

\def\undertext#1{$\underline{\hbox{#1}}$}

\def\pz{{\partial}}

\def\nd{^{\vphantom{\dagger}}}
\def\yd{^\dagger}

\def\vph{\vphantom{\sum_i}}
\def\bvph{\vphantom{\sum_N^N}}

\def\csch{\,{\rm csch\,}}

\def\sgn{\,{\rm sgn\,}}

\def\and{a^{\phantom\dagger}}

\gdef\journal#1, #2, #3, 1#4#5#6{
{\sl #1~}{\bf #2}, #3 (1#4#5#6)}

\def\prb{\journal Phys. Rev. B, }

\def\prl{\journal Phys. Rev. Lett., }

\def\ijmp{\journal Int. J. Mod. Phys., }
\def\jmp{\journal J. Math. Phys., }

\def\jpc{\journal J. Phys. C, }

\def\epl{\journal Europhys. Lett., }
\begin{document}
\draft
\def\uar{\uparrow}
\def\dar{\downarrow}
\def\mmz{{[m_\uar m_\dar 0]}}
\def\vhat{{\hat v}}
\def\rhobar{{\overline\rho}}
\def\sbar{{\overline s}}
\def\fifth{\frac{1}{5}}
\def\uu{{\uar\uar}}
\def\dd{{\dar\dar}}
\def\ud{{\uar\dar}}
\def\du{{\dar\uar}}
\def\izppi{\int\limits_{0^+}^{\infty}}
\def\IB{I_{\rm B}}
\def\gtw{{\tilde{g}}}
\def\ttw{{\tilde{t}}}
\def\thtw{{\tilde{\theta}}}
\def\tot{\frac{3}{2}}
\def\kF{{k_\ssr{F}}}
\def\kJ{{k_\ssr{J}}}
\def\thb{{\bar \theta}}
\def\phb{{\bar \phi}}
\def\Gb{{\overline G}}
\def\II#1#2{I_{#1}^{\ssr[#2\ssr]}}
\def\avg#1{{\overline{[#1]}}}
\def\Var{{\rm Var}}

%\preprint{8-24-93}
\twocolumn[\hsize\textwidth\columnwidth\hsize\csname
@twocolumnfalse\endcsname
\title{Non-linear $I(V)$ Characteristics of Luttinger Liquids and
Gated Hall Bars}

\author{
Scot R. Renn \\
Daniel P. Arovas}
\address{
Department of Physics, University of California at San Diego,
La Jolla, CA 92093}

\date{\today}

\maketitle

\begin{abstract}
Non-linear current voltage characteristics of a disordered
Luttinger liquid are calculated using a perturbative formalism. One
finds
non-universal power law characteristics of the form $I(V)\sim
V^{1/(2\tilde{g}-1)}$ which is
valid both in the superfluid phase when $I$ is small and also in
the insulator phase when  $I$ is large. Mesoscopic voltage
fluctuations are also calculated.  One finds $\Var(\Delta V) \sim
I^{4\tilde{g}-3}$. Both the $I(V)$ characteristic and the voltage
fluctuations exhibit universal power law behavior at the superfluid
insulator transition where $\tilde{g}=\tot$.
The possible application of these results to the non-linear transport
properties of gated Hall bars is discussed.
\end{abstract}

\vskip2pc]

\narrowtext

Recently there has been considerable interest in the transport
properties of the Luttinger liquid\cite{LL}. There are many reasons
for this including applications to quantum wires,  quasi-one
dimensional
organic conductors\cite{Jerome} like TTF-TCNQ and possibly to high
temperature superconductors\cite{Anderson}.
Nevertheless, until  the  recent   observation of Luttinger liquid
behavior by Milliken \etal\ \cite{Milliken}
clear  evidence of Luttinger liquid behavior had been lacking.
In their experiments ,  the point contact  tunneling conductance
between two $\nu=1/3$ fractional quantum Hall edge channels is
measured as a function of the point contact gate voltage.  One
observes transmission resonances whose half-widths scale with
temperature as  $T^{2/3}$. This is in agreement with Kane and
Fisher \cite{Kane}.  Off resonance, the conductance
scales as $T^4$ as was also predicted \cite{Kane}.

In fact, the experiments of  Milliken \etal\ \cite{Milliken}  is only
the most recent contribution
to the understanding of Luttinger liquid transport.  Let us begin our
discussion by recounting,
in a semi-historical manner, some of the major developments in the
field.  We will begin by recalling
the work by Apel and Rice\cite{Apel}, who used the Kubo formalism to
show that the conductance of a clean quantum wire differed from the
usual Landauer result that $G=e^2/h$ \cite{rolf}.  Instead these
authors found that $G=g e^2/h$, where $g\equiv\pi \hbar
v (\partial n/\partial \mu)$.
This result may be obtained as follows: First consider the current
injected into the wire from the left reservoir. This is
$\II{+}{1}= \half ev(\partial n/\partial \mu) \mu_1=g(e/h)\mu_1$.
Similarly the current injected
into the wire from the right reservoir is $\II{+}{2}=
\half ev(\partial n/\partial \mu) \mu_2=g(e/h)\mu_2$.  Now the
total current in the wire is $I\equiv (\II{+}{1}-\II{-}{1})$, where
$\II{-}{1}$ is the current flowing out of the wire into the left
reservoir.
Eliminating $\II{-}{1}$ in favor of $\II{+}{2}$ gives
\begin{equation}
I=g {e\over h}(\mu_1 -\mu_2 ) -\IB
\label{IvsIB}\end{equation}
where $\IB(I)=\II{+}{1}-\II{+}{2}=\II{-}{2} -\II{-}{1}$ is the
backscattering current.  Now  $\IB=0$ for a clean wire, so
eq. \ref{IvsIB} gives $G=I/e(\mu_1-\mu_2)=ge^2/h$.

Another major advance in transport theory
was the investigation of  the superfluid-insulator transition in
uniformly disordered Luttinger liquids by Giamarchi and
Schultz\cite{Giamarchi}. These authors found that an infinitesimal
amount of disorder will localize a Luttinger liquid with $g<\tot$.
Moreover, they find that, if $g>\tot$,  disorder is irrelevant  and
the Luttinger liquid exhibits superfluid behavior.

Further progress  was made by
Kane and Fisher\cite{Kane} who  investigated  a Luttinger wire with
a single or double barrier.
These authors found that $V_{2\kF}$, the Fourier coefficient of the
potential barrier, is highly relevant if $g<1$. Because of this, the
conductance of a Luttinger liquid with  a single impurity vanishes.
The existence of  conductance resonances  may be  crudely understood
if one views Luttinger liquids as
Wigner crystals with quasi-long range positional
order\cite{commentc}. According to this point
of view, the coupling of a CDW to a barrier
potential may be described by the pinning potential
$V_{\rm pin}(\theta)=|V(2\kF)| \cos(2\kF\theta)$
Hence, unless $V(2\kF)=0$,  $G=0$ since the potential barrier will
pin the Wigner crystal.  However, when $V_{2\kF}=0$,
a conductance resonance will occur.

In this paper, we examine the non-linear $I(V)$ characteristic of a
uniformly disordered Luttinger liquid.
Using both renormalization group theory and perturbation theory,
we will analyze the $I(V)$ characteristics   several analytically
tractable regimes. These regimes include the superfluid phase in the
$I\rightarrow 0$ limit and the insulator phase in the
$I\rightarrow \infty$ limit. We will find that, in both instances,
the
$I(V)$ characteristic  is proportional to some
non-universal power of the voltage, $V^{1/(2g-1)}$. However, at the
metal-insulator transition,  one obtains the universal power law
dependence $I\propto V^{1/2}$.
We will also briefly discuss the low voltage behavior of the $I(V)$
characteristic in the insulator phase.

In addition to the general discussion of non-linear transport in
Luttinger liquids, we will also
discuss the  application of these ideas to gated Hall bars.  Here we
consider a device similar to that discussed by Haug \etal\
\cite{Haug}
and  by Washburn\etal\ \cite{Washburn}. The device consists of a Hall
bar divided in half by a long but narrow gate (see Fig. 1).  By
negatively biasing the gate, one can deplete the 2-D electron gas
underneath the gate. In particular, at a critical gate voltage
$V_\rmG$
a pair of edge channels underneath
the gate  will  delocalize so as to allow charge transport  parallel
to the gate.  We  will  argue  that edge channel localization in this
device is an example of the Giamarchi-Schultz superfluid-insulator
transition.

The organization of this paper is as follows:  In section  2, we
introduce the bosonized formulation of the Luttinger liquid and
establish
notation. In section 3, we discuss the
renormalization group calculation of non-linear voltage
characteristics in both the superfluid and insulator phases.
In section 4,  we  discuss the application of these results to gated
Hall bars.

\section{Model}
\label{model}

In this section, we wish to  review the bosonic formulation of
a  uniformly disordered  spinless Luttinger liquid.  This well known
formulation involves the two angle fields $\theta$ and $\phi$ which
obey the commutation relations
\begin{eqnarray}
[\theta(x),\phi(x')]&=&- i\pi\sgn(x-x')
\label{commutator}\end{eqnarray}
 $\theta$ and $\phi$ are related to the number and current
densities associated with the right and left moving
electrons according to
\begin{eqnarray}
\rho_\pm(x)&=&{\nu\over 4\pi}\cbl{\pz\theta\over\pz x}\pm{2\over\nu}
{\pz\phi\over\pz x}\cbr\\{} \cr
j_\pm(x)&=&-{\nu\over 4\pi}\cbl{\pz\theta\over\pz t}\pm{2\over\nu}
{\pz\phi\over\pz t}\cbr\\{}
\end{eqnarray}
The total charge and current densities are
\begin{equation}
\rho= {\nu\over 2\pi}{\partial\theta\over \partial x } \qquad\qquad
j=-{\nu\over 2\pi}{\partial\theta\over\partial t }
\label{rhoj}
\end{equation}
which are identical to the charge and currents of a one-dimensional
CDW provided that $\theta$ is related to the displacement $u$
according
to $\theta(x)=2\kF(x+u(x))$.

The parameter $\nu$ depends on the system under consideration.
For many one dimensional systems, \eg\  Hubbard models,  $\nu=1$.
However, in the context of the theory of fractional
quantum Hall edge states $\nu=1/m$  where $1/m$ is the filling
fraction of the (parent) fractional
quantum Hall liquid \cite{principal}.
Finally, we note the creation
operators of right and left moving  electrons  may be  approximately
written  in the form
\begin{equation}
\psi_\pm\yd(x,t) =\cC \exp(\pm i\theta(x,t)/2)\exp(i\phi(x,t)/\nu)
\end{equation}
where $\cC\simeq 1/\sqrt{2\pi a}$ is a constant determined by the
cutoff structure of the theory \cite{LL} ($a$ is the microscopic
lattice constant).  The angle field $\theta(x,t)$ is separated into
three
parts:$\theta(x,t)\equiv\theta_0(t)+2\kF x +\thtw(x,t)$.  The first
two
terms describe a spatially uniform current and a time-independent
charge
density (cancelled by the background) through eq. \ref{rhoj}.  The
fluctuating piece $\thtw(x,t)$ then describes the quantized
excitations of the
Luttinger liquid.

In the absence of disorder, the Luttinger liquid Hamiltonian is taken
to be
\begin{equation}
\cH_0={\hbar v\over 4\pi}\int\!dx\,\cbl
{1\over 2\gtw}\({\pz\thtw\over\pz x}\)^2
+ 2\gtw\({\pz\phi\over\pz x}\)^2-{2\nu e\over\hbar v} E\,\thtw\cbr
\label{Ho}\end{equation}
where $E(x)$ is the electric field, and $v$ is the charge velocity.
However, in order to describe dirty Luttinger liquids, we must
include the effect of point impurities,
located at positions $x_i$, into our Hamiltonian.
This is done by including a backscattering contribution  $\cH_\rmB$
to our total Hamiltonian
$\cH=\cH_0+\cH_\rmB$. $\cH_\rmB$ is given by
\begin{eqnarray}
\cH_{\rm B}&=&-\sum_i \left[t\nd_i\, \psi\yd_+(x_i)\psi\nd_-(x_i)
+ t^*_i\, \psi\yd_-(x_i)\psi\nd_+(x_i)\right].\nonumber\\
&=& -{1\over 2\pi a}\sum_i  \left[t\nd_i\,e^{i\theta_0}\,e^{2i\kF
x_i}\,e^{i\thtw(x_i)} +{\rm H.c.}\right]
\label{HB}
\end{eqnarray}
and is time-dependent through $\theta_0(t)$.
The backscattering amplitudes $t_i$ are assumed to be uncorrelated,
\ie
\begin{equation}
\langle t\nd_i t_j^*\rangle = |t|^2\,\delta_{ij}\ .
\label{BS1}
\end{equation}
For future reference, we introduce
\begin{equation}
D\equiv n\nd_{\rm imp}|t|^2 a/2\pi v^2 \hbar^2,
\label{BS2}
\end{equation}
a dimensionless measure of disorder.

In the following sections, we will  obtain non-linear $I(V)$
characteristics by calculating the backscattering current which is
defined by $\IB=\int dx\> {i}_\ssr{B}(x)$, where  $\dot{\rho}_+(x) +
\partial_x \rho_+(x)=- {i}_\ssr{B}(x)$.  For steady states this
implies that $\IB =I^1_+-I^2_+$  the definition introduced after eq.
\ref{IvsIB}.
One can readily obtain an explicit expression for ${\rm \IB}$  (and
${i}_\ssr{B}$) by comparing the Heisenberg equations of motion with
$\dot{\rho}_+(x) + \partial_x \rho_+(x)=- {i}_\ssr{B}(x)$.
The result is
\begin{eqnarray}
\IB&=&{i\over 2\pi a\hbar}\sum_i\left[t_i\,e^{i\theta_0}\,e^{2i\kF
x_i}\, e^{i\thtw(x_i)}- {\rm H.c.} \right]
\label{IBdef}\end{eqnarray}
Since the backscattering operator $\psi\yd_+(x)\psi\nd_-(x)$, which
transfers an electron from the $-$ channel to the $+$ channel,
is proportional to
$e^{i\thtw(x)}$, much of our analysis is based on the behavior of the
correlation function
\begin{equation}
\xhi(x,t)={i\over\hbar}
\langle[e^{i\thtw(x,t)},e^{-i\thtw(0,0)}]\rangle\,\Theta(t)
\label{CF}
\end{equation}
which has been studied by Luther and Peschel \cite{corr}.  A
discussion of its
behavior can be found in Appendix A.

\section{Superfluid-Insulator Transition}
\label{super}

A renormalization group treatment of the above
model has been given by Giamarchi and Schultz \cite{Giamarchi}.
They find that when the microscopic cutoff length $a\to a\, e^l$,
that the coupling constants flow
to $\gtw(l)$, $ D(l)$, $v(l)$ and $E(l)$ which obey the flow
equations
\begin{eqnarray}
\bvph{d\gtw(l)\over dl}&=&-\gtw(l)^2D(l) \cr
\bvph{dD(l)\over dl} &=&-2(\gtw(l) -\tot)D(l) \cr
\bvph{dv(l)\over dl} &=&-\gtw(l)\,v(l)\,D(l) \cr
\bvph{dE(l)\over dl} &=&0
\end{eqnarray}
which are illustrated in Fig. 2.  For $E=0$, there exists
a superfluid phase which is the domain of attraction of a line of
fixed points and an insulator phase.  The phase boundary is at
$\Delta=0$
where $\Delta^2 =\frac{9}{4}D-(\gtw-\tot)^2$.
This is a Kosterlitz-Thouless transition since, near the
$(D,\gtw)=(0,\tot)$
fixed point, the RG flow equations are equivalent to the Kosterlitz
Thouless RG equations. Because of this, a variety of well known
results
apply to this model. For instance, in the insulating state, near
the MI transition, the
localization length is simply
$\xi_L=\exp(\pi/\Delta)=\exp(\pi/\sqrt{D-D_c})$.
This has the same form as the correlation length of the $XY$ model.
Now invariance under the renormalization group implies that
$I(E;X,a)=I(E;X(l),ae^l)$. A simple rescaling of $(x,t)\to
(x\,e^{-l},t\,e^{-l})$ will, according to dimensional analysis, give
\begin{eqnarray}
I(E;X,a)=e^{-l}I(e^{2l}E;X(l),a)
\label{scaling}
\end{eqnarray}
This general result will be used in the next section to
check perturbative calculation of $I(V)$.

\section{The Non-Linear $I(V)$ Characteristic of Dirty Luttinger
Liquids}

In this section, we wish to perturbatively
calculate the non-linear $I(V)$ characteristic of a Luttinger liquid.
The result gives non-universal power law $I(V)$
characteristics.  Using  eq. \ref{scaling}, we will
argue that the perturbative $I(V)$ characteristics
are valid  both in the superfluid phase when $I\rightarrow 0$ and
in the insulator phase when $I\rightarrow \infty$.

Our approach to obtaining  $I(V)$ begins with a calculation of
$\IB(V)$ to order  $\cO(|t|^2)$ in perturbation theory.
To do this, we assume that a steady-state current $I$ flows along the
wire.
This is described by taking $\theta_0(t)=\Omega t$,
with $\Omega=2\pi I/\nu e$.
Now according to linear response theory  the backscattering current
to leading order in $t_i$ is
\begin{equation}
\IB=-{ie\over\hbar}|\cC|^4\sum_{i,j}\left[t\nd_i
t^*_j\,e^{iG(x_i-x_j)}
\xhi(x_i-x_j,\Omega)-{\rm c.c.}\right]
\label{IBeqn}
\end{equation}
where $\xhi(x,t)$ is the response function discussed in Appendix A,
and where $G=2\kF+d/\ell^2$.  If we now average over the impurity
backscattering
amplitudes as in eqs. \ref{BS1}-\ref{BS2}, we find
\begin{eqnarray}
\vph\overline{[\IB]}&=&{2e\over\hbar}\,|\cC|^4\,n\nd_{\rm imp}|t|^2
L\, \xhi''(x=0,\Omega)\label{IB}\\
&=&\bvph{D\over\Gamma(2\gtw)} \({L\over a}\)\({ev\over a}\)
\({2\pi a|I|\over\nu e v}\)^{2\gtw-1}\sgn I\quad (T=0)\cr
&\simeq& C(\gtw) \,{2\pi D\over\nu}\({L\over a}\)
\({\pi\gtw\kT a\over\hbar v}\)^{2\gtw-2} I
\qquad(T\gg {hI\over\nu e k_\ssr{B}})\nonumber
\end{eqnarray}
where $\overline{[\IB]}$ denotes an average with respect to the
impurity
positions, and $C(\gtw)$ is a numerical factor (see Appendix A).

To determine when this is valid, we calculated the higher order
terms in perturbation theory.  We obtain, at $T=0$,
\begin{equation}
{[\IB]\over L}= \sum_{n=1}^{\infty}  a_n D^n
I^{(2\gtw-1)+(2\gtw-3)(n-1)}
\label{series}
\end{equation}
where $a_n$ are numerical coefficients. The above result implies
that higher order corrections to eq. \ref{IB} are negligible if
$|D(Ia/ev)^{2\gtw-1}|<<1$.
This means that  eq. \ref{IB}  is valid in  the superfluid phase for
small $I$ and is valid in insulator phase for large $I$.  These
conclusions may also be reached from
renormalization group arguments. However,  these arguments
indicate that the observed exponent depends not on the bare $\gtw$
but rather on the $\gtw_*$ the governing fixed point.  Depending on
whether
$I\rightarrow 0$ or $I\rightarrow \infty$ limit is taken, the fixed
point may be an  infrared or ultraviolet fixed point.

Now using eq. \ref{IB} and eq. \ref{IvsIB} one can solve for
$I(\mu_1-\mu_2)$.
In the limit where $L\rightarrow \infty$  but $I$ is held fixed, one
obtains a non-linear $I(V)$ characteristic of the form
\begin{equation}
I \propto {ev\over a}\({eE a^2\over \hbar v D}\)^{1/(2\gtw_*-1)}
\label{IvsE}
\end{equation}
There are several comments to be made about the above result. The
first is that the linear response regime was
not considered since it automatically breaks down
in the $L\rightarrow \infty$ limit. The next comment is that, when
$D<<\Delta$, the above result is consistent with the scaling identity
derived in eq. \ref{scaling}.  The verification is straightforward,
one simply demonstrates that
$e^{-l}( e^{2l}E/D(l))^{1/(2\gtw-1)}$  is $l$-independent.
For example, when  $\gtw>\tot$, one uses
$D(l)=4\Delta^2/(9 \sinh^2(\Delta l +b_0)) \sim e^{-2\Delta l}$ where
$\Delta \approx(\tot -\gtw)$ to verify the $l$-independence.
$\gtw<\tot$ may similarly be demonstrated
provided that $l$ is  not so large that $D(l)<<\Delta$ is violated.
The  last comment regarding eq. \ref{IvsE} is that  the insulating
phase can display a negative differential resistance depending on
whether
$\gtw_*<1/2$ or $\gtw_*>1/2$ , eq. \ref{IvsE}.

At this point, we wish to discuss briefly the issue of mesoscopic
fluctuations.
This is a very interesting topic which will be discussed at length
elsewhere\cite{inprogress}. However, here we simply wish to
understand when $\Var(\Delta \mu)<< |\Delta \mu |$ which the
condition which must be met in order that the  predicted $I(V)$
characteristic will not be obscured by
mesoscopic voltage fluctuations. To do this, we first calculate
$\Var(\IB)$ at $T=0$ using eq. \ref{IBdef}. One finds that
\begin{equation}
{\Var(\IB)\over L}=\({\hbar ev^2 D\over\pi a^3}\)^2 \int\!
{dq\over 2\pi}\>[\xhi'' (2\kF,\Omega)]^2
\label{VarIB}\end{equation}
where $\xhi(q,\omega)$ is the Fourier transform of $\xhi(x,t)$.
Then, using $\Delta\mu\equiv\mu_1-\mu_2=h \IB/\nu^2 e\gtw$, and
evaluating
eq. \ref{VarIB}, we find
\begin{equation}
{\Var(\Delta \mu)\over L}={16 \pi\over a}
{\Gamma(2\gtw-1)^2\over\gtw^2\Gamma(\gtw)^2\Gamma(4\gtw-2)}
\({Dvh\over\nu^2 a}\)^2\({2\pi I a\over\nu e v}\)^{4\gtw-3}\ .
\end{equation}
This implies
\begin{equation}
{\Var(\Delta \mu)\over [\Delta \mu]^2}=
{\pi\Gamma(2\gtw-1)^2\Gamma(2\gtw)^2\over\Gamma(\gtw)^2
\Gamma(4\gtw-2)}\, {\xi_I\over L}
\label{vardmu}
\end{equation}
where $\xi_I \equiv \nu ev/2\pi I$ is a length scale associated
with the finite current. Mesoscopic voltage fluctuations will be
small if $L\gg\xi_I$.

\section{Devices}
\label{devices}

The experimental work of Milliken \etal\ \cite{Milliken}, has raised
the hope of using
gated semiconductor heterostructure devices as convenient laboratory
models of the Luttinger liquid.  It is therefore of interest to
consider devices which might be used to experimentally investigate
dirty
Luttinger
liquids.  Of course, the main requirement of such a device is that it
allows backscattering to occur at random points, uniformly
distributed
along a finite length of  a pair of oppositely directed  but parallel
edge channels.  This geometry may be  achieved  in several devices.
 One device is a
mesoscopic Hall bar of width $w$ of order of a few magnetic lengths.
This sort of device has been studied by Simmons \etal\
\cite{Shayegan} in
the context of measuring the quasi-particle charge.

Another interesting device is the gated Hall bar (see Fig. 1).
In this second device\cite{comment}, one has a long but very narrow
gate (i.e. gate width $d\sim \cO(\ell)$ where $\ell =\sqrt{\hbar
c/eB}$ is the magnetic length).  We will suppose that one has an
incompressible $\nu=1/m$
fractional quantum Hall liquid and that the gate is straddled
by only  a single pair of edge channels. The two pair of edge
channels will contribute
the left and right movers of a chiral  Luttinger liquid model
equivalent to that discussed in
section 1. To see this we follow Wen\cite{WenCDW,Wenreview} and write
down a simple edge Hamiltonian
\begin{equation}
\cH_0=\half \int \! dx\ \left\{U_\ssr{RR} [\rho^2_\ssr{R} +
\rho^2_\ssr{L}]
+2U_\ssr{RL}[\rho_\ssr{R} \rho_\ssr{L}] \right\}
\end{equation}
where $[\rho_{\alpha}(x),\rho_{\beta}(x')]=\mp (\nu/2 \pi)
\delta_{\alpha\beta}\partial_x \delta(x-x')$.
If we identify $\rho_\ssr{L}+\rho_\ssr{R}=-\nu \partial_x
\theta/2\pi$ and
$\rho_\ssr{L} -\rho_\ssr{R} =
-\partial_x \phi/\pi$, then $H_0$  reduces to the model defined in
eqn. \ref{Ho}-\ref{HB} and the commutators reduce to that given in
eqn. \ref{commutator} where
\begin{equation}
g=\nu\left[{U_\ssr{RR}-U_\ssr{RL}\over
U_\ssr{RR}+U_\ssr{RL}}\right]^{1/2}
=\nu^2\,\gtw
\end{equation}

Next we consider interchannel tunneling, the analogue of
backscattering.
Provided that one neglects  inter-Landau level mixing associated with
the gate potential,
then it is well known that a tunneling event  involves a
change of momentum equal to $\Delta k= d/\ell^2$. Hence, tunneling is
suppressed between a pair of infinite rectilinear edges.
More realistically, however, the
the gate potential will not be uniform in systems with such extremely
narrow gates. Instead it will
exhibit random spatial irregularities and as a result the edges
follow the irregular equipotential contours of Fig. 3.  In this case,
interchannel tunneling  will occur  particularly  near the points of
close contact.  The points of close contact in Fig. 3 are denoted
with dots.
Now  we observe that, since the path of each chiral edge is not
straight, the distance traveled between  a pair of adjacent
impurities is
different for the two edges. One must therefore write the physical
tunneling operator as
$t_i \psi\yd_+(x_i -\half s_i)\psi\nd_-(x_i +\half s_i) + {\rm H.c.}$
where
$x_i-x_{i-1}\mp  (s_i -s_{i-1})$ is the
path length between the $i^{\rm th}$ and $(i-1)^{\rm th}$ impurities
on the $\pm$ edge channels.

Now expanding the bosonized tunneling operator in $s_i$ and
dropping gradient terms in $\theta$ gives
\begin{eqnarray}
\cH_\rmt&\approx&-\!\!\int\!\!dx\, \left[\ttw(x)\, e^{idx/\ell^2}\,
e^{i\theta(x)}+{\rm H.c.}\right]
\nonumber\\
\ttw(x)&=& {1\over 2\pi a}\sum_i \ttw_i\,\delta(x-x_i)\nonumber
\end{eqnarray}
We observe that whether or not  the phase of $t_i$ is random,
the phase of $\ttw_i=t_i\,e^{is_i d/\ell^2}$ will be random provided
that $\Var(s_i)\gg(4\pi \ell^2/d)^2$.
If this is the case, then our model reduces to that discussed in
section 1.
We note that the model does not include quasiparticle tunneling which
is suppressed by the $\nu=0$ depletion zone separating the edge
channels.

Consider now the behavior of $(\gtw,D)$ as a function of the gate
voltage in this device.
When the gate has a large negative bias, the two edges will be well
separated, backscattering will be suppressed, and
$(\gtw,D)\rightarrow (1/\nu, 0)$. This puts the system on
the superfluid side of the phase diagram.
The edge channels straddling the gate can then readily transport
charge across the Hall bar.  Now suppose that the $|V_\rmG|$
decreases.
As this occurs, the edges get closer together. This increases $D$ and
decreases $\gtw$.  The
antiwire Hamiltonian, therefore, moves across the phase diagram on
the
dotted line as shown schematically in Fig. 2. At some gate voltage
$V^*$, a superfluid-insulator transition occurs. For smaller
$|V_\rmG|$ the
antiwire is localized.

What is the manifestation of the superfluid insulator transition
in Hall bar?  To understand this, we need to know how $I(V)$
characteristics of the
two terminal antiwire are related to the four terminal
characteristics of the gated
Hall bar.  Qualitatively, the longitudinal voltage drop $\mu_6-\mu_5$
{\it vs.}\ source-drain current $I_{14}$ is similar to the $I(V)$
characteristic of the two terminal antiwire. The reasons for this are
twofold.
First, the source-drain current $I_{14}$ gives rise to a voltage
drop across the antiwire. Secondly the antiwire current $I_\rmA$
produces a longitudinal voltage $\mu_6 -\mu_5$ in the Hall bar.

Below we will derive the actual  quantitative relation between
Hall bar and antiwire characteristics.  We will find that
\begin{eqnarray}
\bvph\mu_6 -\mu_5 &=&{h\over\nu e} F[(\mu_6 -\mu_5)/e +
(h/\nu e^2)I_{14}] \cr
\bvph\mu_2-\mu_6 &=&{h\over\nu e} I_{14}
\label{Deltamu}
\end{eqnarray}
In the above equations,
$I_\rmA=F(V_\rmA)$ is the non-linear characteristic of a segmented
wire which we will construct from  the edge channels straddling the
gate and from the edges
which connect the antiwire to terminals \#2, 3, 5, and 6.
These connecting channels are paired up, as shown in Fig. 4, in order
to give two non-chiral Luttinger liquids connected to the two ends of
the antiwire.

The three pieces of the segmented wire, labeled 1, 2, and 3, are
characterized by different $g$'s. In particular, $g_1=g_3=\nu$.
However, the renormalized $g_2\equiv\nu^2\gtw$ of the disordered
central segment depends on the interaction
strengths $U_\ssr{RR}$, and $U_\ssr{RL}$, $\nu$, and
the  disorder $D$. Now because $g$  at the end segments is not equal
to that in the interior segment, eq. \ref{IvsIB} is modified to
\begin{equation}
I_\rmA=F(V_\rmA)= \nu {e^2\over h}V_\rmA -\IB(I_\rmA, \gtw)
\label{char}
\end{equation}
where $\IB(I_\rmA,\gtw)$ is the backscattering current which
occurs only in segment B.  One remarkable feature of this result
is that, in the absence of backscattering,
the conductance of the segmented antiwire $G_\rmA\equiv
I_\rmA/V_\rmA$
is independent of $g_2$.

At this point, we  need to introduce the notation
$\II\pm\alpha$  (where $\alpha=1,\dots,6$)  which denotes the
chiral edge current entering ($-$) or leaving ($+$) reservoir
$\alpha$. We should note that the
chiral currents are excess currents defined relative to
a reference state in which the source-drain current $I_{14}$
vanishes.  The sign convention for these currents is as follows:
$\II{\pm}{\alpha}$ is taken to be positive if the {\it excess}
current has the same sense as the propagation direction of the
capillary waves \ie\ it is in the direction of the arrows decorating
the edge channels in Fig. 1.

We can now derive eqn. \ref{Deltamu}. First we must enumerate the
current conservation conditions:
The first of these are the four conditions of the form
$\II{+}{1}=\II{-}{2}$, $\II{+}{3}=\II{-}{4}$,
$\II{+}{4}=\II{-}{5}$, $\II{+}{6}=\II{-}{1}$.
In addition, there are 6  additional equations associated with
current conservation at the terminals.  These are
$\II{+}{1}-\II{-}{1}=I_{14}$, $\II{-}{4}-\II{+}{4}=I_{14}$, and
$\II{+}{\alpha}=\II{-}{\alpha}$
for $\alpha=2,3,5,6$.  Finally, there is a single constraint
associated with current conservation across
the antiwire. This is $\II{+}{2}-\II{-}{6}=\II{-}{3} -\II{+}{5}$.

Now in order to obtain 12 equations in the 12 unknown
$\II{\pm}{\alpha}$,
we need to augment the current conservation equations with
the antiwire characteristic:
\begin{equation}
\II{+}{2}-\II{-}{3}=F[(\nu e^2/h)(\II{+}{2} -\II{+}{5})]
\end{equation}
The form of the antiwire characteristic  follows
from the fact that the voltage across the antiwire is
$e\Delta V_\rmA=(\mu_5 -\mu_2)=(h/\nu e)(\II{+}{2} -\II{+}{5})$,
whereas the
antiwire current is $I_\rmA=\II{+}{2}-\II{-}{3}$.

The twelve equations are readily solved to give $\II{\pm}{\alpha}$ in
terms of the source-drain current $I_{14}$.  From this,
one then calculates the voltage drop across an arbitrary pair of
terminals using
$\II{+}{\alpha}=(\nu e/h)\mu_{\alpha}$\cite{derivation}.
In this manner, eq. \ref{Deltamu} is obtained.

Notice that eq. \ref{Deltamu} says that the antiwire current is
\begin{displaymath}
I_\rmA={\nu e\over h}(\mu_6-\mu_5)\ .
\end{displaymath}
This result, together with eq. \ref{char}, gives us the $I(V)$
characteristic
\begin{displaymath}
I_{14}=\IB(I_\rmA)=\IB\Bigl((\nu e/h)(\mu_6-\mu_5)\Bigr)\ ,
\end{displaymath}
which, through eq. \ref{IB}, may be written in a form appropriate for
the gated Hall bar:
\begin{displaymath}
I_{14}=I_0\({L\over L_D}\)
\({\mu_6-\mu_5\over e V_0}\)^{2\gtw-1}\ ,
\end{displaymath}
where $V_0\equiv\hbar v/ea$, $I_0\equiv\nu ev/2\pi a$,
and $L_D\equiv \nu\Gamma(2\gtw)\,a/2\pi D$.
The above non-universal power law is valid in the superfluid phase
when
$\mu_6-\mu_5$ is small, and in the insulator phase when $\mu_6-\mu_5$
is
large.  At the superfluid-insulator transition, the above result
gives
a universal power law behavior of the form
$I_{14}/L\propto(\mu_6-\mu_5)^2$.

In the superfluid phase,
the above $I(V)$ characteristic should be observable throughout a
window
$I_{\rm min}\ll I_{14}\ll I_{\rm max}$ whose upper limit is set
by the applicability of perturbation theory, and whose lower limit
is set by the condition that the results not be obscured by
mesoscopic
fluctuations.  From Appendix B, we see that higher order terms in
perturbation
theory come in powers of $D (I/I_0)^{2\gtw-3}$, hence a rough
estimate
for $I_{\rm max}$ is $I_{\rm max}\approx I_0 D^{-1/(2\gtw-3)}$.
{}From eq. \ref{vardmu}, the condition
$\avg{I_{14}^2}_\rmc/\avg{I_{14}}^2
\approx 1$ gives us $I_{\rm min}\approx I_0 (L/L_D) (A_\gtw
a/L)^{2\gtw-1}$,
where $A_\gtw$ is a known numerical factor.
We assume that the magnetic length $\ell$ provides a rough lower
bound to the ultraviolet cutoff~$a$; at $B=10\,\rmT$, then,
$a\gtwid\ell=81\AA$. We also assume $\nu=\third$, so that
$\gtw=3$ in the limit of widely separated
edge channels (a wide barrier), and a capillary wave velocity of
$v=10^6\ \rmm/\rms$.  Now $D\propto |t|^2$ is a Gaussian function of
the distance between the two edge channels and can therefore be
arbitrarily small;
at the superfluid-insulator transition, $D\sim\cO(1)$.  Thus
$I_0\approx 1\ \mu\rmA$, $V_0\approx 80\ \rmm\rmV$, and
$L_D\approx (500/D)$~\AA.  For an antiwire of length $L=20\ \mu\rmm$,
then, we estimate $I_{\rm min}\approx (8D)\ \rmn\rmA$ and $I_{\rm
max}\approx
D^{-1/3}\ \mu\rmA$.  The window $I_{\rm min}\ll I \ll I_{\rm max}$
is larger for wider barriers.

Although at the time of writing, there have been several experiments
involving gated hall bars\cite{Haug,Washburn}, no experiments
studying either the superfluid insulator transition or the non-linear
transport properties of a Luttinger liquid have been attempted.
Clearly an experimental investigation into either of these aspects of
Luttinger liquid physics would be most welcome.

\section{Acknowledgements}

The authors gratefully acknowledge conversations with R.C. Dynes, S.
Applebaum, S. Coppersmith, S. Kivelson, G. Zimanyi, and M. P. A.
Fisher. This research was supported in part by
NSF grant DMR-8957993 (DPA), by
a fellowship from the Alfred P. Sloan Foundation (SRR),
and by the Donors of the Petroleum Research Fund, administered by the
American Chemical Society.

%\vfill\eject

\bigskip

\centerline{\uppercase
{\bf Appendix A: Correlations at}}
\centerline{\uppercase
{\bf Finite Temperature}}

\smallskip

The correlation function
\begin{displaymath}
\xhi(x,t)={i\over\hbar}\langle[e^{i\thtw(x,t)},e^{-i\thtw(0,0)}]
\rangle \, \Theta(t)
\end{displaymath}
has been calculated by Luther and Peschel \cite{corr}.  In real
space,
one finds
\begin{displaymath}
\xhi(x,t)={2i\over\hbar}\,\Theta(t)\,{\rm Im}\left[
J_\gtw(vt-x) J_\gtw(vt+x)\right]
\end{displaymath}
where
\begin{displaymath}
J_\gamma(z)=\left[{a-iz\over a}{\hbar v\over\pi z\kT}
\sinh\({\pi z\kT\over\hbar v}\)\right]^{-\gamma}\ .
\end{displaymath}
Here $T$ is the temperature and $a$ is a short distance cutoff.
The Fourier transform $\xhi(q,\omega)$ is then
\begin{eqnarray*}
\xhi(q,\omega)&=&{1\over 4\pi\hbar
v}\int\limits_{-\infty}^\infty\!\!\!
{ds\over s-i 0^+}\,\Bigl[K_\gtw(s+\omega-vq)\,K_\gtw(s+\omega+vq)\\
&&\qquad-K_\gtw(-s-\omega+vq)\,K_\gtw(-s-\omega-vq)\Bigr]\\
\bvph K_\gamma(\eta)&=&\int_{-\infty}^\infty\!\!\!\!\!du\
\,e^{iu\eta}\>
\({a\over a+iu}\)^\gamma\({\pi\kT u\over\hbar v}\,
\csch{\pi\kT u\over\hbar v}\)^\gamma\\
&&\simeq \Gamma(1-\gamma)\,{\rm Im}\,\(-\eta-i\,
{\pi\gamma\kT\over 2\hbar v}\)^{\gamma-1}\ ,
\end{eqnarray*}
where the last line is an approximation given by Luther and Peschel.
For our purposes, we need $\xhi''(q,\omega)$, which at $T=0$ is
\cite{corr}
\begin{displaymath}
\xhi''(q,\omega)={\pi^2 a^{2\gtw}\over\Gamma^2(\gtw)\,\hbar v}
\({\omega^2-v^2 q^2\over 4v^2}\)^{\gtw-1}\Theta(\omega^2-v^2
q^2)\,\sgn(\omega)
\end{displaymath}
and at high temperatures given by
\begin{displaymath}
\xhi''(q,\omega)\simeq C(\gtw)\, {a^{2\gtw}\over\hbar v}\,
\({\omega\over 2v}\)
\({\pi\gtw\kT\over 2\hbar v}\)^{2\gtw-3}\,\Theta(\pi\gtw\kT-\hbar
v|q|)
\end{displaymath}
where $C(\gtw)$ is independent of $\omega$, $q$, and $T$.
We also need the $\xhi(x=0,\omega)$, the integral
of $\xhi(q,\omega)$ over all $q$.  This is given by
\begin{displaymath}
\xhi''(x=0,\omega)={\pi\, a^{2\gtw}\over\Gamma(2\gtw)\,\hbar v}\,
\left|{\omega\over v}\right|^{2\gtw-1}\sgn(\omega)
\end{displaymath}
at low temperatures and
\begin{displaymath}
\xhi''(x=0,\omega)\simeq C(\gtw)\,{a^{2\gtw}\over\hbar v}\,
\({\omega\over v}\) \({\pi\gtw\kT\over\hbar v}\)^{2\gtw-2}
\end{displaymath}
at high temperatures.

\bigskip
\centerline{\uppercase
{\bf Appendix B: Dimensional Analysis of}}
\centerline{\uppercase
{\bf Perturbation Series}}
\smallskip

Here we discuss the derivation of Eq.(\ref{series}), in which it was
claimed that the perturbative result for $\IB$ is
\begin{equation}
{\IB\over L}= \sum_{n=1}^{\infty}  a_n D^n I^{(2g-3)n+2}\ .
\end{equation}
This says that the higher order terms in the series are small for
large $I$
if $g<\frac{3}{2}$, and for small $I$ if $g>\frac{3}{2}$.

To derive this result, we begin with the Euclidean action
$S=S_0+S_1$,
\begin{eqnarray*}
S_0 &=& {1\over 8\pi v\gtw}\int\!dx\,d\tau
\left[\(\pz\thtw\over\pz\tau\)^2
+v^2\(\pz\thtw\over\pz x\)^2\ \right] \cr
S_1 &=& \int\!dx\,d\tau \left[ e^{-i\Omega\tau}\,\ttw(x)\,
e^{-i\thtw(x,\tau)}
+e^{i\Omega\tau}\,\ttw^*(x)\,e^{i\thtw(x,\tau)} \right]\cr
Z&=&Z[\ttw(x,\tau),\ttw^*(x,\tau)]\equiv\int
\rmD\thtw\,e^{-(S_0+S_1)}\ ,
\end{eqnarray*}
where $\ttw(x,\tau)\equiv \ttw(x)e^{-i\Omega\tau}$, and consider the
Green's function
\begin{eqnarray}
\lefteqn{\Bigl\langle\exp i\sum_{k=1}^n\(\thtw(x_k^+,\tau_k^+)-
\thtw(x_k^-,\tau_k^-)\)\Bigr\rangle_0=e^{-nG(0)}\qquad\qquad\vph}
\nonumber\\
& &\qquad\times\prod_{i,j}
|z_i^+ -z_j^-|^{-2\gtw}\Bigg/\prod_{i<j}|z_i^+ -z_j^+|^{-2\gtw}
|z_i^- -z_j^-|^{-2\gtw}
\label{green}
\end{eqnarray}
where $z_j^\pm \equiv x_j^\pm + i\tau_j^\pm$ are the complexified
space-time
coordinates.   Here, $G(x,\tau)\simeq 2\gtw\ln(R/|z|)$ is
proportional to
the Green's function for the Laplacian in a two-dimensional disk of
size  $R$.  The limit $G(0)$ is rendered finite by a suitable
ultraviolet
cutoff.  Now the numerator of eq. \ref{green} is homogeneous and of
degree $-2n^2\gtw$, while the denominator is homogeneous and of
degree
$-2n(n-1)\gtw$.  A general term of order $|\ttw|^{2n}$ in the
perturbation
expansion for $\langle\IB(\tau)\rangle$ will involve an
integral over $2n$ space coordinates $\{x_1^\pm,\ldots,x_n^\pm\}$ and
$2n-1$ time coordinates $\{\tau_1^+,\tau_2^\pm,\ldots,\tau_n^\pm\}$.
Clearly the integral is invariant
under an overall spatial translation $x_k^\pm\to x_k^\pm + \Delta x$,
so the
$4n-1$ integrations produce an overall factor of the length $L$ and
result in a degree of homogeneity of $4n-2$.  Note that since
$\tau\equiv\tau_1^-$
is not integrated, there is no overall time translational invariance.
Finally, the disorder average pairs each $x_j^+$ with some $x_k^-$,
resulting in a loss of $n$ spatial integration variables.  The
overall degree of homogeneity is then
\begin{eqnarray*}
{\rm deg}(n)&=&(4n-2)-n-2n^2\gtw+2n(n-1)\gtw\\
&=&(3-2\gtw)n -2\ .
\end{eqnarray*}
Thus, assuming that the ultraviolet divergences are cancelled
(as in the linear response case), we conclude that the
$n^{\rm th}$ order term in the perturbation expansion behaves as
\begin{equation}
\IB^{(n)}\propto \Omega^{2+(2\gtw-3)n}\propto
I^{(2\gtw-1)+(2\gtw-3)(n-1)}
\end{equation}
which is what we set out to show.

\vspace{0.4 in}

\undertext{FIGURE CAPTIONS}

\vspace{0.1 in}

FIG.1   A Hall bar with a  long but narrow  gate. (The gate width  $w
\sim \cO(\ell)$).  Terminals \#1 and \#4 are the source and drain
terminals,
respectively.  The Hall bar is decorated with the edge state geometry
discussed in the text. The two parallel edge segments straddling the
gate
are referred to as the antiwire in the text.  As discussed in the
text, the
antiwire provides an experimental realization of a dirty Luttinger
liquid.

FIG. 2  The renormalization group flows of a dirty spinless Luttinger
liquid.
The superfluid and insulator phases are separated by the line
$D^{1/2}=2\gtw/3 -1$.  The dashed line
indicated the trajectory of an antiwire system as a function of gate
voltage.

FIG. 3  A schematic drawing of a pair of edge channels under an
ultranarrow gate.  Because of the irregular nature of the
equipotentials, the
depletion gap between the edge channels
exhibits spatially random fluctuations.  Electron tunneling across
the $\nu=0$ depletion gap is assumed to occur at points of close
contact
which are indicated in the figure with dots.

FIG. 4 The tri-segmented antiwire which is constructed from the pair
of edge channels straddling the gate plus the four edge channels
which
connect the gate to terminals \#2, \#3, \#5, and \#6.  The connecting
edge
channels to terminals \#2 and \#3 are paired to give segment 1.
The edge channels straddling the gate form segment 2, and the
connecting edge channels connecting the antiwire
to terminals \#5 and \#6 form segment 3. Note that $g_1=g_3=\nu$,
but because of disorder and interaction effects, $g_2\neq\nu$.


\begin{references}
\bibitem{principal} We consider only simple quantum Hall edge states,
arising
from so-called ``principal'' bulk QHE fluids, and not the
composite edges which occur, for example, at filling fraction
$\nu=2/3$.
\bibitem{comment} This is essentially identical with devices studied
by Haug \etal\ and also by Washburn \etal
\end{references}
\end{document}